\documentclass[11pt]{article}

\usepackage{amsmath} 
\usepackage{amssymb}
\usepackage{amsthm} 
\usepackage{bbm}
\usepackage{bm} 
\usepackage{graphicx}
\usepackage{cite}
\usepackage{pifont}
\usepackage[makeroom]{cancel}
\usepackage[usenames,dvipsnames]{color}
\usepackage[colorlinks=true,citecolor=blue,linkcolor=blue,urlcolor=blue]{hyperref}
\usepackage{braket}

\usepackage{todonotes}

\newcommand{\de}{\partial}
\newcommand{\lr}[1]{\left(#1\right)}

\def\Re{{\rm Re}}
\def\Im{{\rm Im}}
\def\k{{\bf k}}
\def\x{{\bf x}}
\def\y{{\bf y}}

\begin{document}
\title{\LARGE\textbf{Causality and classical dispersion relations}}
\author{Raphael E. Hoult, Pavel Kovtun\\
\vspace{-3mm}\\
{\it\small Department of Physics \& Astronomy, University of Victoria}\\
{\it\small PO Box 1700 STN CSC, Victoria, BC, V8W 2Y2, Canada}\\
\vspace{-3mm}\\
}

\date{September, 2023}

\maketitle

\begin{abstract}
\noindent 
We explore the consequences of relativistic causality and covariant stability for short-wavelength dispersion relations in classical systems. For excitations described by a finite number of partial differential equations, as is the case in relativistic hydrodynamics, we give causality and covariant stability constraints on the excitation's frequency at large momenta. 
\end{abstract}

\section{Introduction and summary}
We are interested in the following question: for a relativistic physical system which admits small excitations whose frequency $\omega$ is related to the wave vector $\k$ by the dispersion relation $\omega(\k)$, what do causality and stability imply about the functional form of $\omega(\k)$? 
Let us start with (linear) stability. 
This notion is relatively straightforward to define in ``classical'' theories, by which we mean theories in which the signals in question are described by a finite number of partial differential equations. Linearizing the equations about a given solution, the excitations of the classical theory are the eigenmodes of the linearized partial differential equations. Linearizing homogeneous equations about a constant solution representing an equilibrium/ground state, and taking the perturbations proportional to $\exp(-i\omega t + i\k{\cdot}\x)$ will give rise to the dispersion relations $\omega(\k)$. One example is Maxwell's equations in matter, where $\omega(\k)$ describe electromagnetic waves. In fluid-dynamical equations, $\omega(\k)$ describe mechanical and thermal perturbations of the fluid, such as sound waves and shear waves. Stability of these linearized perturbations implies that $\Im(\omega(\k))\leq0$ for real $\k$; the inequality corresponds to dissipation of small excitations. 

We next define the following notion of ``classical causality'':
\begin{align}
\label{classical-causality}
  \parbox{0.8\textwidth}{\em A classical theory is causal if its equations are (at least weakly) hyperbolic, and normals to the characteristics lie outside the light-cone.}
\end{align}
Hyperbolicity gives rise to a finite propagation speed, and characteristic normals outside the light-cone ensure that the propagation speed is below the speed of light (for an elaboration refer to Appendix \ref{ap:real_space}). The requirement of classical causality%
\footnote{
  It is worth emphasizing that weak hyperbolicity of the equations, while necessary in order to define classical causality, does not guarantee local well-posedness. 
}
will constrain  the possible forms of $\omega(\k)$ which can emerge from a causal (according to the above definition) classical theory. Here we will use the term ``classical theory'' to denote a description of a $D$-dimensional physical system in terms of a finite number of partial differential equations for functions of $D$ variables.

In quantum relativistic theory, on the other hand, causality is defined through the vanishing of (anti-)commutators of local operators outside the light cone~\cite{Streater-Wightman}. Among other things, this ``microscopic causality'' has implications for retarded Green's functions. As a simple example, consider two local bosonic operators $A(x)$ and $B(x)$, whose retarded Green's function is $G^R(x-y) = i\theta(x^0 - y^0) \langle [A(x), B(y)]\rangle$, where $\theta(z)$ is the step function, and angular brackets denote the expectation value in the equilibrium/ground state.  As the commutator vanishes outside the light-cone, microscopic causality implies that the retarded Green's function is not only proportional to $\theta(x^0 - y^0)$, but also proportional to $\theta\left(c^2(x^0 {-} y^0)^2 - (\x {-} \y)^2 \right)$, where $c$ is the speed of light. Fourier transforming $G^R(x-y)$, one arrives at $\tilde G^R(\omega,\k)$.  The first theta-function implies that, when $\k$ is real, $\tilde G^R(\omega,\k)$ is an analytic function of $\omega$ in the upper complex half-plane, $\Im(\omega)>0$. Taken together, the two theta-functions imply%
\footnote{
  Under the standard assumption of no exponential growth in spacetime.
}
that $\tilde G^R(\omega,\k)$ is analytic when $\Im(\omega) > c \, |\Im(\k)|$. 

Thus, the Fourier transform of the retarded function can only have singularities when
\begin{align}
\label{eq:GFC}
  \Im(\omega) \leq c\,|\Im(\k)| \,.
\end{align}
While the causality condition~\eqref{eq:GFC} is textbook material in relativistic quantum theory~\cite{Itzykson:1980rh}, its implications for {\em classical} relativistic theories appear somewhat under-explored. Suppose that the macroscopic dynamics of quantities corresponding to $A$ and $B$ is described by a classical theory in the appropriate macroscopic limit (as an example, $A$ could be the energy density, and $B$ the particle number density). Linear response then dictates that the eigenmodes of the classical theory appear as singularities of $\tilde G^R$, in other words $\tilde G^R(\omega(\k), \k)^{-1} = 0$. See for example~\cite{Forster} for a discussion. Thus the dispersion relations $\omega(\k)$ of the classical theory must satisfy \eqref{eq:GFC}, in order to ensure that the classical theory is consistent with microscopic causality. 

Recently, refs.~\cite{Heller:2022ejw, Heller:2023jtd} applied the inequality \eqref{eq:GFC} to power-series expansions of $\omega(\k)$ at small $\k$ in relativistic theories (including relativistic hydrodynamics), and derived constraints on hydrodynamic transport coefficients in terms of the convergence radius of the expansion of the function $\omega(\k)$ about $\k=0$. The goal of the present paper is to explore the consequences of the constraint \eqref{eq:GFC} at large~$\k$, where the notion of causality is conventionally defined in classical theories~\cite{Krotscheck1978}. From now on, we will set the speed of light $c$ to one. 

The constraint \eqref{eq:GFC} on the eigenfrequencies of a classical system may be equivalently viewed in terms of stability: if the system is stable in all reference frames, then  \eqref{eq:GFC} follows \cite{Gavassino:2021owo, Gavassino:2023myj}. Thus, one can also refer to the microscopic causality constraint \eqref{eq:GFC} as a covariant stability constraint.  

In classical theories, if one is interested in $\omega(\k)$, there is more to the causality conditions than what one can naively read off from eq.~\eqref{eq:GFC}.  As an example, consider the equation
\begin{equation}
\label{eq:bad_wave_eq}
  (\partial_t^2 \partial_x^2 - \partial_x^4) \varphi(t,x) = 0.
\end{equation}
Its eigenfrequencies $\omega(k) = \pm k$ satisfy \eqref{eq:GFC}, however the equation itself is inconsistent with classical causality: in fact, \eqref{eq:bad_wave_eq} is not even hyperbolic, let alone causal. If, on the other hand, one is interested in working with $\k(\omega)$ instead of $\omega(\k)$, then, the eigenmomenta are $k(\omega)=\pm\omega$, and $k(\omega)=0$; the latter clearly does not satisfy  \eqref{eq:GFC}, hence eq.~\eqref{eq:bad_wave_eq} is not covariantly stable. 

The constraint \eqref{eq:GFC} in fact contains much more than just the requirement of sub-luminal propagation.  It is possible for a classical system to be sub-luminal, while at the same time violating \eqref{eq:GFC} due to instabilities. As an example, consider the equation
\begin{align}
\label{eq:ude}
  \left(\partial_t^2 - \partial_t - \partial_x^2 \right) \varphi(t,x) = 0.
\end{align}
The equation is causal yet unstable, and its eigenfrequencies $\omega(k)$ do not satisfy \eqref{eq:GFC}. 

For classical linear perturbations of a given equilibrium/ground state, causality as defined by \eqref{classical-causality} is a significantly weaker statement than the covariant stability condition \eqref{eq:GFC}. Every linear classical theory that is covariantly stable is causal, but, as the previous example illustrates, not every causal theory is covariantly stable. 

Let us now come to the consequences of the covariant stability condition \eqref{eq:GFC} for large-$\k$ dispersion relations in classical theories. For real $\k$, covariant stability implies 
\begin{align}
\label{eq:c-1}
  0 \leq \lim_{|\k| \to \infty} \frac{| \Re \left( \omega(\k) \right)|}{|\k|} \leq 1\,, \ \ \ \ 
  \lim_{|\k| \to \infty} \frac{\Im\left( \omega(\k) \right)}{|\k|} = 0\,.
\end{align}
In a given classical theory, such as relativistic fluid dynamics, the dispersion relations $\omega(\k)$ are determined as solutions to $F(\omega,\k) = 0$,
where $F(\omega,\k)$ is a polynomial of finite degree in $\omega$ and $\k$ whose exact form is determined by the differential equations of the classical theory. One can show that classical causality as defined in \eqref{classical-causality} amounts to three conditions on $\omega(\k)$. The first two conditions are given by eq.~\eqref{eq:c-1}, while the third condition concerns the number of ``modes'', i.e.\ the number of solutions when $F(\omega,\k) = 0$ is solved for $\omega$ at fixed non-zero $\k$. 

This third condition is
\begin{align}
\label{eq:c-2}
  {\cal O}_{\omega}\biggl[F(\omega, \k\neq 0)\biggr] = {\cal O}_{|\k|}\biggl[F(\omega = a |\k|, \k = {\bf s} |\k|)\biggr] \,,
\end{align}
where $a$ is a non-zero real constant,  ${\bf s}$ is a real unit vector, and ${\cal O}_{z}$ denotes the order of the polynomial in the variable $z$. A linear classical theory whose dispersion relations $\omega(\k)$ satisfy the conditions \eqref{eq:c-1}, \eqref{eq:c-2} is causal. In our earlier example \eqref{eq:bad_wave_eq}, we had $F(\omega, k) = k^2(\omega^2 - k^2)$; hence, the theory is not causal because of its violation of \eqref{eq:c-2}, even though it respects \eqref{eq:c-1}.

In a rotation-invariant theory, in a rotation-invariant state, choose $\k$ along $\x$, and let $k\equiv k_x$. For large $k$ (not necessarily real), dispersion relations in a classical covariantly stable theory admit convergent expansions as $|k| \to\infty$:
\begin{align}
\label{eq:wk-2}
  \omega(k) = \sum_{n=0}^{n_0} c_{1-2n} k^{1-2n} +  c_{-2n_0} k^{-2n_0} + ... \,.
\end{align}
The leading-order coefficient $c_1$ is real with $0 \leq |c_1|\leq 1$, and $n_0$ is a non-negative integer. All of the coefficients $c_{1-2n}$ are real, and could be zero. The term $c_{-2 n_0}$ must have $\Im(c_{-2n_0}) \leq 0$. The dots refer to subleading terms, which do not necessarily come with integer powers of $1/k$.

Our work is motivated by causal theories of relativistic hydrodynamics. As we will later explain, causal theories of relativistic hydrodynamics must contain non-hydrodynamic modes. 
Even though the existence of such modes in classical theories is required by causality, the physics described by such causality-restoring modes is outside of the validity regime of the hydrodynamic approximation, as has been appreciated for many years, see e.g.~\cite{Geroch:1995bx}. This is not surprising, given that causality restoration corresponds to large-$\k$ physics, unlike hydrodynamic excitations which describe small-$\k$ physics. In general, one may hope that choosing the classical causality-restoring modes in a way that mimics the true large-$\k$ behavior of the fundamental microscopic theory will improve the predictive power of the classical theory. Clearly, the ability of classical theories to mimic the large-$\k$ behavior of an interacting quantum field theory is limited at best. Our results may be viewed as a way to quantify precisely what the limitations of classical theories are when mimicking the true large-$\k$ physics, and what is beyond their abilities.

The constraints stated above are quite elementary to derive, and we do so in the next section. Following the derivation, we will discuss a few illustrative examples of the above causality conditions.  Appendix \ref{ap:real_space} shows that the conditions \eqref{eq:c-1} and \eqref{eq:c-2} can be viewed as a restatement of classical causality \eqref{classical-causality} for linear partial differential equations. In Appendix \ref{ap:ProofPowerSeries}, we prove that classical systems admit convergent expansions in the $|\k| \to \infty$ limit, and that the first term in these expansions is always integer. 

{\em Note added:} As we were preparing the final version of the manuscript, we received a preliminary version of the preprint~\cite{Wang:2023csj}, whose results have overlap with ours, and appears on arXiv on the same day.

\section{Implications of covariant stability}

We consider a classical system in $D$-dimensional flat space, described by partial differential equations for functions of $D$ variables $x^\mu$. The equations are taken to be Lorentz covariant, so that Lorentz transforms of solutions solve Lorentz transformed equations. In order to find the eigenmodes of the classical system, one takes the unknown functions $U^A(x)$ as $U^A(x) = \bar U^A + \delta U^A(x)$, where $\bar U^A$ are constant solutions, representing the ground/equilibrium state of interest, and linearizes the equations in $\delta U$. The resulting linear system of differential equations, $L_{AB}[\partial]\delta U^B = 0$, can be solved by the Fourier transform, $\delta U^A(x) = \delta \tilde U^A(K) \exp(i K_\mu x^\mu)$, where $K_\mu = (-\omega, \k)$. Non-trivial solutions exist provided 
\begin{align}
\label{eq:spectral_curve}
  F(\omega,\k) \equiv \det L[i K] = 0\,.
\end{align}
Solving this equation gives rise to the dispersion relations $\omega(\k)$. Let us now explore the consequences of eqs.~\eqref{eq:GFC} and \eqref{eq:spectral_curve} for the large-$\k$ dispersion relations. 

For simplicity, we investigate the case where the linearized equations are rotation-invariant. An example of such a system is given by relativistic fluid dynamics, when the background solution $\bar U^A$ describes a fluid at rest. We choose $\k$ along $\x$, and let $k\equiv k_x$. The spectral curve $F(\omega,k)$ is a finite-order polynomial in $\omega$ and $k$. The dispersion relations $\omega(k)$ are determined by the polynomial equation $F(\omega,k)=0$, which describes a complex curve in ${\bf C}^2$. We are interested in the solutions $\omega(k)$ of this polynomial equation, in the limit $k\to\infty$. The Puiseux theorem \cite{Wall-singular-points} implies that the solutions $\omega_i(k)$ can be expanded in convergent Laurent series as $k\to\infty$, 
\begin{align}
\label{eq:w-pe}
  \omega(k) = \sum_{m=m_0}^\infty c_{m} \zeta^m \,,
\end{align}
where $\zeta^r = 1/k$, and $r$ is a positive integer. If $M$ is the order of the polynomial $F(\omega,k)$ as a function of $\omega$, then there are $M$ expansions \eqref{eq:w-pe} for each solution $\omega(k)$. The modes come as $N$ sets, each with $r_a$ branches, such that $\sum_{a=1}^N r_a = M$. In particular, each term in the expansion \eqref{eq:w-pe} is $c_m (e^{2\pi i \ell/r})^m /k^{m/r}$, where $\ell=0, 1,  \dots ,r{-}1$ for the $r$ branches. 

Let us look at the leading-order term in this expansion, $\omega(k) \sim C k^p + \dots$, where $p$ is a real rational exponent, and the coefficient $C$ is in general complex, $C = C' + i C''$. For complex momentum $k = \kappa e^{i\alpha}$, covariant stability \eqref{eq:GFC} implies
\begin{align}\label{eq:GFC_expansion}
  \kappa^{p-1} \left( C'' \cos(p\alpha) + C' \sin(p\alpha) \right) \leq |\sin \alpha| \,.
\end{align}
Taking $\alpha=0$, $\alpha=\pi/p$, and $\alpha= \pm\pi/2p$ in this equation implies that the dispersion relations are at most linear, $\omega(k\to\infty) \sim C k$, where the coefficient $C$ is real, with $|C|\leq1$.%
\footnote{
  If $\omega(k\to\infty) \sim C k$ with $-1\leq C\leq1$ in one reference frame, it remains so in all reference frames. 
}
Therefore, in the expansion \eqref{eq:w-pe}, we must have $m_0\geq-r$ for all modes. 
Supposing the linear term is non-zero, depending on the value of $r$ for a given set of modes, the expansion thus may proceed in the following way: 
\begin{align}
  r&=1: &  \omega(k) & = c_{-1} k + c_0 + c_1/k + \dots\,, \\
  r&=2: & \omega(k) & =  c_{-2} k + c_{-1} k^{1/2} + c_0 + c_1/k^{1/2} + \dots\,, \\
  r&=3: & \omega(k) & = c_{-3} k + c_{-2} k^{2/3} + c_{-1} k^{1/3} + c_0 + c_1/k^{1/3} + \dots\,,
\end{align}
In general, since the linear term is real for $\alpha=0, \pm\pi$, the next term in the expansion can also be constrained by \eqref{eq:GFC_expansion}, though in a more limited fashion. Setting $p$ to be less than unity, one can see that if $p$ is non-integer (and therefore $r>1$), the next term will generically violate \eqref{eq:GFC_expansion} due to the phase factor in \eqref{eq:w-pe}. The next term must therefore come with an integer power of $1/k$; moreover, its coefficient must be such that $\Im(c_n) \leq 0$, as may be seen by setting $\alpha = 0$ in \eqref{eq:GFC_expansion}. If the next term after the linear term has an odd-integer power of $1/k$, the coefficient must be real, $\Im(c_n) = 0$, as may be seen by setting $\alpha = \pm \pi$.

Terms with real coefficients do not appear in \eqref{eq:GFC_expansion} when $\alpha = 0, \pm \pi$, and so if the next term after the leading-order term is odd (and therefore has a real coefficient), condition \eqref{eq:GFC_expansion} also constrains the \textit{next} term after the next-to-leading-order term. If that term is even, then it must have negative imaginary part, and \eqref{eq:GFC_expansion} does not (immediately) constrain the following sub-leading terms, including possible fractional terms. If it is real, the process repeats.

Therefore, non-integer terms may only begin appearing after the first even-integer term in the expansion. The expansion must generically be of the form
\begin{equation}\label{eq:expansion_generic}
    \omega(k) = \sum_{n=0}^{n_0} c_{1-2n} k^{1-2n} +  c_{-2n_0} k^{-2n_0} + ... \,,
\end{equation}
where $n_0$ is a non-negative integer, the $c_{1-2n}$ are all real, and any (or all) of the $c_{1-2n}$ may be zero. Additionally, $|c_1| \leq 1$, and $\Im(c_{-2 n_0}) \leq 0$. The dots denote higher-order subleading terms, which may include fractional powers of $1/k$. 

Some examples of expansions which are \textit{not} ruled out by eq.~\eqref{eq:GFC_expansion} are the following:
\begin{align}
\label{eq:wkfrac1}
    \omega(k) &= c_{1} k + c_{0}  + c_{-1/2} k^{-1/2} + \dots\,,\\
    \omega(k) &= c_{0} + c_{-1/2} k^{-1/2} + \dots\,,\\
    \omega(k) &= c_{1} k + c_{-1} k^{-1} + c_{-3} k^{-3} + c_{-4} k^{-4} + c_{-9/2} k^{-9/2}+\dots\,.
\end{align}
It may be possible to extract covariant stability constraints on the terms beyond the first even-integer term. We plan to return to exploring further large-$k$ constraints in the future. 

Returning to linear order, one may straightforwardly show that condition \eqref{eq:c-2} follows from \eqref{eq:GFC} as well. Suppose one had a polynomial spectral curve $F(\omega,k)$ which did \textit{not} satisfy condition \eqref{eq:c-2}. Then the highest order terms in the polynomial (giving $\omega$ and $k$ the same power counting, as $\omega$ may be at most linear in large-$k$ by conditions \eqref{eq:c-1}) must necessarily have an overall factor of $k$:
\begin{equation*}
    F(\omega, k) = k^{g-M} G(\omega,k) + ... = 0 \,,
\end{equation*}
where $M$ is the order of the polynomial $G$ in $\omega$, and $...$ denotes lower-order terms. Suppose one then Lorentz-boosted the system longitudinally in $\omega, k$. Then $k^{g-M} \to \lr{k' - v \omega'}^{g-M}$, where $-1\leq v\leq 1$ is the boost parameter. There are therefore $g-M$ new modes which appear at linear order in large-$k$ which violate conditions \eqref{eq:c-1}, and therefore \eqref{eq:GFC}.

Another condition may be extracted by making use of condition \eqref{eq:c-2}. This condition ensures that the spectral curve $F(\omega,k)$ is of the form
\begin{equation}
    F(\omega,k) = a_M(k) \omega^M + a_{M-1}(k) \omega^{M-1} + ... + a_{1}(k) \omega + a_{0}(k)
\end{equation}
where the $a_{N}(k)$ are polynomials in $k$ of order $\ell_N \leq M-N$. One can show then that unless all $a_N(k)$ are $k$-independent, there must be at least one branch of the large-$k$ expansion of $\omega(k)$ which has a non-zero linear term. Refer to appendix \ref{ap:ProofPowerSeries} for more details.

\section{Examples and discussion}
\label{sec:discussion}
As a simple example, consider the diffusion equation 
\begin{align}
\label{eq:diffusion-1}
  (\partial_t - D \partial^i \partial_i) \varphi(t,\x) = 0\,,
\end{align}
where $D>0$ is the diffusion constant, and $\varphi$ is a scalar field. The corresponding dispersion relation $\omega(\k) = -i D \k^2$ violates the second condition in \eqref{eq:c-1}, hence the diffusion equation is not causal. A violation of causality implies a violation of covariant stability~\cite{Gavassino:2021owo}, hence the relativistic covariant version of the equation must be unstable. The covariant equation is 
\begin{align}
\label{eq:diffusion-2}
    (u^\mu \partial_\mu - D \Delta^{\mu\nu}\partial_\mu \partial_\nu) \varphi(t,\x) = 0\,,
\end{align}
where the unit timelike velocity vector $u^\mu$ specifies the rest frame of the diffusing matter, $\Delta^{\mu\nu} \equiv g^{\mu\nu} + u^\mu u^\nu$ is the spatial projector, and $g^{\mu\nu}$ is the inverse (flat-space) metric. At small $\k$, there is indeed an instability due to a mode which behaves as $\omega(\k\to0) =  i/(D \gamma v^2)$, where $v$ is the spatial velocity of $u^\mu$, and $\gamma = (1-v^2)^{-1/2}$ is the relativistic boost factor. Similarly, choosing the boost velocity along $\k$, one finds $\omega(k\to\infty) = k/v + \dots$, with an acausal leading order term, and an unstable subleading term. See e.g.\ ref.~\cite{Kostadt:2000ty} for a discussion of eq.~\eqref{eq:diffusion-2}.

As another example, consider modifying the diffusion equation by a higher-derivative term:
\begin{align}
\label{eq:diffusion-3}
  \left( \partial_t - D\, \partial^i \partial_i - \tau D\, \partial_t \partial^i \partial_i \right) \varphi(t,\x) = 0\,,
\end{align}
where constant ``relaxation time'' $\tau>0$. The dispersion relation is $\omega(\k) = -i D \k^2/(1+\tau D \k^2)$, interpolating between diffusive behavior $\omega = -i D \k^2 + \dots$ at small $\k$, and a constant value $\omega = -i/\tau + \dots$ at large $\k$. Even though $\omega(\k)$ obeys both of the conditions \eqref{eq:c-1}, eq.~\eqref{eq:diffusion-3} is not causal because the third condition \eqref{eq:c-2} is not obeyed. The covariant equation is
\begin{align}
\label{eq:diffusion-4}
    (u^\mu \partial_\mu - D (1+\tau u^\alpha \partial_\alpha) \Delta^{\mu\nu}\partial_\mu \partial_\nu ) \varphi(t,\x) = 0\,.
\end{align}
Choosing the boost velocity $v$ along $\k$, one finds modes which behave as $\omega(k\to\infty) = k/v + \dots$, with an acausal leading order term, and an unstable subleading term. Alternatively, the dispersion relation $\omega(\k) = -i D \k^2/(1+\tau D \k^2)$ has a simple pole at $\k^2 = -1/(\tau D)$, while, as emphasized in \cite{Krotscheck1978, Heller:2022ejw}, simple poles in dispersion relations are forbidden by causality. See also ref.~\cite{Gavassino:2023mad} for related comments. 

As another example, consider a hyperbolic version of the diffusion equation, sometimes called the telegraph equation, 
\begin{align}
\label{eq:te}
   \left( \frac{c_s^2}{D} \partial_t - c_s^2 \partial^i \partial_i  +  \partial_t^2 \right) \varphi(t,\x) = 0\,.
\end{align}
Here $0\leq c_s \leq 1$ determines the wave front speed, and $D$ is the diffusion constant. The dispersion relation at short wavelength, $\omega(\k) = \pm c_s |\k| + \dots$, is causal. The mode counting condition \eqref{eq:c-2} is obeyed as well, hence eq.~\eqref{eq:te} is causal. The modes are also stable for all $\k$ (for positive $D$), hence the covariant equation 
\begin{align}
\label{eq:te-2}
   \left( \frac{c_s^2}{D} u^\mu \partial_\mu - c_s^2 \Delta^{\mu\nu} \partial_\mu \partial_\nu  +  u^\mu u^\nu \partial_\mu \partial_\nu \right) \varphi(t,\x) = 0\,.
\end{align}
is stable, and its dispersion relations $\omega(\k)$ satisfy eq.~\eqref{eq:GFC}. At negative $D$, the theory would be causal but unstable. 

As another example, consider dispersion relations determined by $F(w, q)=0$, where 
\begin{align}
  F(w, q) = 8 q^2 - 16 i w - 8 i q^2 w - 16 w^2 + 4 i w^3 + (q^2 - w^2)^2\,,
\end{align}
where $w\equiv \omega \Gamma$, $q\equiv k \Gamma$, in terms of some dimensionful parameter $\Gamma$. 
At small $k$, there is a diffusive mode, and three stable gapped modes. The dispersion relations satisfy both \eqref{eq:c-1} and \eqref{eq:c-2}, hence this linear theory is causal. However, this theory is unstable at {\em large} $k$, because there are modes for which $\lim_{k\to\infty} \Im(\omega) = +\infty$.

As a stable example, consider dispersion relations determined by $F(w, q)=0$, where 
\begin{align}
\label{eq:diff-cs}
  F(w, q) = 4 q^2 - 4 i w - 4 i w q^2 - 8 w^2 + 4 i w^3 + (q^2 - w^2)^2\,,
\end{align}
where again $w\equiv \omega \Gamma$, $q\equiv k \Gamma$, in terms of a dimensionful parameter $\Gamma$. At small $k$, there is a diffusive mode $\omega(k) = -i \Gamma k^2 + O(k^4)$, and three gapped eigenfrequencies with negative imaginary parts: one gapped mode is purely imaginary, and two gapped modes are off the imaginary axis. At large $k$, the eigenfrequencies are 
\begin{align}
  \omega(k) = \pm k - \frac{i}{\Gamma}  \pm \frac{1{-}i}{\Gamma} \frac{1}{(2\Gamma k)^{1/2} } + \dots\,,
\end{align}
providing an example of the expansion \eqref{eq:wkfrac1} in a causal and stable theory. 
The theory \eqref{eq:diff-cs} is stable in all reference frames, and may be viewed as another way to modify the diffusion equation at short distances in a way that preserves causality.

As our next example, consider a classical theory of M\"uller-Israel-Stewart type, applied to hydrodynamics of conformal fluids~\cite{Baier:2007ix}. The polynomial $F(\omega, k)$ which determines the dispersion relations in the rest frame of the fluid is given by $F(\omega, k) = F_{\rm shear}(\omega, k)^2 F_{\rm sound} (\omega, k)$. The shear factor is $F_{\rm shear}(\omega, k) = i\omega - \gamma k^2 + \tau \omega^2$, where $\gamma$ is the diffusion constant for transverse momentum density, and $\tau$ is the stress relaxation time. The sound factor is $F_{\rm sound}(\omega, k) = \omega^2 - \frac13 k^2 + i\omega k^2 (\frac13 \tau + \frac43 \gamma) - i\tau \omega^3$. The shear modes obey the telegraph equation, and the large-$k$ dispersion relations are $\omega_{\rm shear}(k) = \pm \sqrt{\gamma/\tau} \, k + \dots$. For the sound mode, the large-$k$ dispersion relations are $\omega_{\rm sound}(k) = \pm \sqrt{\frac13 + \frac43 \gamma/\tau}\, k + \dots$, and $\omega_{\rm sound} = -i/(\tau+4\gamma) + \dots$. All modes are stable, and causality of the linearized theory is preserved for $\gamma/\tau< 1/2$. The large-$k$ expansions proceed in integer powers of $1/k$. The reason is that the propagation velocities are different for the three modes, hence each mode admits a Puiseux expansion \eqref{eq:w-pe} with $r=1$.

As a final example, let us look at large-$k$ dispersion relations arising from the singularities of retarded functions of the energy-momentum tensor in strongly coupled ${\cal N}=4$ supersymmetric Yang-Mills theory~\cite{Festuccia:2008zx, Fuini:2016qsc}. There are infinitely many modes labeled by integer $n$, whose large-$k$ expansion at real $k$ yields
\begin{align}
\label{eq:wk-sym}
  \omega^{\pm}_n(k) = \pm |k| \pm c_n e^{\mp i \pi/3} \lr{\pi T}^{4/3}|k|^{-1/3} + \dots \,,
\end{align}
where $T$ is temperature, and $c_n$ is a positive constant whose value depends on which components of the energy-momentum tensor give rise to the retarded function in question. While classical causality is consistent with the leading term, covariantly stable classical theories cannot describe the subleading $|k|^{-1/3}$ term because of the fractional power. This is not surprising: the ${\cal N}=4$ supersymmetric Yang-Mills theory is not classical, and the holographic description of this $3+1$ dimensional theory which gives rise to eq.~\eqref{eq:wk-sym} proceeds in terms of partial differential equations for functions of 5 (rather than $4$) variables. In general, in a quantum or statistical theory, $\lim_{k\to\infty} \omega(k)$ may depend on the phase of $k$, and there is no reason for the Puiseux expansion \eqref{eq:w-pe} to apply.%
\footnote{
  As another explicit example, one can consider 2+1 dimensional quantum field theories that are dual to anti-de Sitter gravity in 3+1 dimensions. In such theories, the paper \cite{Grozdanov:2023txs} found that the dispersion relation of the shear mode goes as $\omega_{\rm shear}(k) \sim C k^4$ at large real $k$, with a non-zero constant $C$. On the other hand, at large imaginary $k$, the function $\omega_{\rm shear}(k)$ can not grow faster than $k$ in order to be consistent with covariant stability (assuming $\omega_{\rm shear}$ stays on the positive imaginary axis at purely imaginary $k$, as seems to be the case). Thus, in such holographic theories, the large-$k$ limit of $\omega_{\rm shear}(k)$ depends on the phase of $k$. We thank S.~Grozdanov for bringing Ref.~\cite{Grozdanov:2023txs} to our attention. 
}
Thus no covariantly stable classical hydrodynamic theory in $3+1$ dimensions would be able to mimic the subleading behaviour in the dispersion relations~\eqref{eq:wk-sym}. 

Let us summarize our results. In this paper, we have explored the consequences of the the covariant stability constraint~\eqref{eq:GFC} for large-$\k$ dispersion relations arising in linear classical theories. The necessary and sufficient conditions of causality are given by eqs.~\eqref{eq:c-1} and \eqref{eq:c-2}. 
While the first equation \eqref{eq:c-1} has long been used as a criterion of causality, the second equation in \eqref{eq:c-1} has perhaps not been as appreciated. For example, the standard reference \cite{Krotscheck1978} assumes that $\Im(\omega)$ vanishes at large $k$, which is too restrictive for causal classical systems, and does not hold in causal theories of relativistic hydrodynamics. 
The condition \eqref{eq:c-2}, while simple to state, has not received significant attention in the physics literature. It says that the number of eigenmodes $\omega(\k)$ must be equal to
\begin{align}
\label{eq:c-2a}
  g\equiv {\cal O}_{|\k|}\biggl[F(\omega = a |\k|, \k = {\bf s} |\k|)\biggr]\,,
\end{align}
where ${\cal O}_{|\k|}$ is the order of the polynomial in $|\k|$, and gives the number of modes in the large-$\k$ limit. The number of modes in a system is fixed regardless of the value of $|\k|$, and so if there are $g$ modes in the large-$|\k|$ limit, there are also $g$ modes in the small-$|\k|$ limit. This is another perspective on why, in relativistic viscous hydrodynamics, non-hydrodynamic (gapped) modes are required to ensure causality -- the number of hydrodynamic (gapless) modes is simply not high enough to have a causal theory.

Our final new result is that all terms in the large-$k$ expansion of $\omega(k)$ before the first non-vanishing even-integer term in $(1/k)$ must be integer, and odd, as stated in eq.~\eqref{eq:wk-2}. This constrains the appearance of any terms with non-integer powers of $(1/k)$ to be after the first term with an even-integer power of $(1/k)$. An example of a causal and stable classical theory with fractional powers of $(1/k)$ in the large-$k$ expansion of $\omega(k)$ is given by eq.~\eqref{eq:diff-cs}.
It may be possible to constrain terms beyond the first even-integer term using other methods, something which would present an interesting area for future exploration. 

If a system is shown to be causal and stable in one inertial reference frame, it is causal and stable in all inertial reference frames~\cite{Gavassino:2021owo}. With the constraints \eqref{eq:c-1} and \eqref{eq:c-2} in hand then, there is a simple algorithmic procedure to be enacted in the rest frame of a rotationally-invariant system, which checks whether a given system of linear partial differential equations represents causal dynamics:

{\it i)} Take all fields proportional to  $\exp(-i\omega t + ik x)$, and determine the polynomial $F(\omega, k)$ which gives rise to the dispersion relations; 

{\it ii)} Check whether the order of the polynomial $F(\omega, k)$ in $\omega$ is equal to $g$ in eq.~\eqref{eq:c-2a}; 

{\it iii)}  Find the large-$k$ wave velocities $a$ by solving $\partial^{g}\!/\partial k^g F(\omega = a k, k) = 0$; 

{\it iv)} If all large-$k$ velocities $a$ are real with $-1\leq a\leq 1$, the theory is causal; 

{\it v)} Further, if the roots of $F(\omega,k)=0$ satisfy $\Im(\omega)\leq 0$ for all real $k$, the theory is covariantly stable. 

The important new point in the above procedure is ensuring that the condition in step {\it ii)} is satisfied. 
Point {\it iv)} can be assured by imposing that $\partial^g/\partial k^g F(\omega =  a k, k)$ is a polynomial in $a$ with real roots which obeys Schur's stability criterion. Point {\it v)} can be assured by demanding that $F(\omega = i \Delta, k)$ is a polynomial in $\Delta$ which obeys the Routh-Hurwitz stability criterion.

We hope that the procedure outlined above will be helpful for exploring causal and covariantly stable effective classical descriptions, such as covariantly stable theories of relativistic hydrodynamics. 

\paragraph{Acknowledgements}
This work was supported in part by the Natural Sciences and Engineering Research Council of Canada (NSERC). A part of this work was completed during the program ``The Many Faces of Relativistic Fluid Dynamics'' at the Kavli Institute for Theoretical Physics at UC Santa Barbara. We thank KITP for their hospitality. This work was supported in part by the National Science Foundation under Grant No. NSF PHY-1748958. We would like to thank M.~Disconzi, L.~Gavassino, J.~Noronha, A.~Serantes, and B.~Withers for helpful discussions. We would also  like to thank the authors of \cite{Wang:2023csj} for helpful correspondence.

\appendix


\section{Real-Space Constraints}\label{ap:real_space}
Consider a system of $n$ linear partial differential equations of order $m$ with constant coefficients%
\footnote{
While the following analysis can be repeated for mixed-order systems, it is more complicated. For simplicity, we restrict ourselves to systems of partial differential equations of the same order.
},
\begin{equation}
\label{ap2:linearized}
   {\cal A}^{\mu_1 \mu_2 ... \mu_m}_{AB} \de_{\mu_1} \de_{\mu_2}... \de_{\mu_m} \delta U^B + {\cal B}^{\mu_1 \mu_2 ... \mu_{m-1}}_{AB} \de_{\mu_1} \de_{\mu_2} ... \de_{\mu_{m-1}} \delta U^B + \,\,...\,\, + {\cal C}_{AB} \delta U^B = 0\,.
\end{equation}
The matrices ${\cal A}, {\cal B}, \dots, {\cal C}$ are constant real $n\times n$ matrices, and $\delta U^A(x)$, with $A=1,\dots, n$, are the unknown functions. The causal structure is defined by the flat-space Minkowski metric $\eta_{\mu\nu}$, and the unknown functions $\delta U^A$ transform under representations of the Lorentz group, so that the equations \eqref{ap2:linearized} are Lorentz-covariant.  
The system of partial differential equations is causal if, given initial conditions with compact support, the solution at a later time has compact support only within the causal future of the initially supported region. In other words, this means that the characteristics of the theory (which define the wavefronts of the theory) lie within the lightcone.
The characteristics of the theory are determined by the characteristic equation \cite{Courant-Hilbert},
\begin{equation}
\label{ap2:char_eq}
    Q(\xi_{\mu_1}, \xi_{\mu_2}, ..., \xi_{\mu_m}) \equiv \det\lr{{\cal A}^{\mu_1 \mu_2 ... \mu_m} \xi_{\mu_1} \xi_{\mu_2} ... \xi_{\mu_m}} = 0\,,
\end{equation}
where the covectors $\xi_\mu$ are normal to the characteristics. Alternatively, if characteristics are level-sets of a scalar function $\phi(x)$, then $\xi_\mu = \de_\mu \phi$. Subluminal propagation speeds correspond to the normals $\xi_\mu = (\xi_0, \bm{\xi})$ pointing outside the lightcone, i.e.\ $\eta^{\mu\nu} \xi_\mu \xi_\nu \geq 0$, i.e.\ the $\xi_\mu$ satisfying the characteristic equation~\eqref{ap2:char_eq} must be spacelike.  Therefore, in a given reference frame, one can find the solutions to the characteristic equation of the form $\xi_0 = \xi_0(\bm{\xi})$, and impose the following constraints on these solutions:
\begin{align}
\label{eq:xi-12}
\frac{|\Re(\xi_0(\bm{\xi}))|}{|\bm{\xi}|} \leq 1, \qquad \Im(\xi_0(\bm{\xi})) = 0 \,.
\end{align}
The first condition imposes that $\xi_\mu$ is spacelike; the second condition demands that the characteristics be real, and therefore the system is not elliptic. Once eqs.~\eqref{eq:xi-12} are true in a given reference frame, they will of course continue to hold in all reference frames. 
However, in a given reference frame, it may so happen that there are solutions to the characteristic equation \eqref{ap2:char_eq} which are not of the form $\xi_0 = \xi_0(\bm{\xi})$. 
Any solution that cannot be written as $\xi_0 = \xi_0(\bm{\xi})$ is necessarily of the form $f({\bm \xi}) = 0$, for all~$\xi_0$. Such solutions to the characteristic equation do not constrain $\xi_0$, which can be  arbitrarily large. Therefore, characteristics of this type will stray outside the lightcone, and a classical theory in which the characteristic equation~\eqref{ap2:char_eq}, in a given reference frame, contains a $\xi_0$-independent factor $f(\bm{\xi})$ will violate causality. A simple way to eliminate such acausal theories is to impose a condition on the number of solutions to the characteristic equation that are of the form $\xi_0 = \xi_0(\bm{\xi})$,
\begin{equation}
\label{eq:xi-3}
    \text{num}(\xi_0(\bm{\xi})) = {\cal O}_{|\bm{\xi}|} \left[ Q(\xi_0 = a |\bm{\xi}|, \bm{\xi} = \bm{s} |\bm{\xi}|) \right] \equiv g \,, 
\end{equation}
where $a$ is an arbitrary real constant, $\bm{s}$ is a unit vector, and ${\cal O}_{|\bm{\xi}|}$ denotes the (maximum) order of the polynomial in $|\bm{\xi}|$. 
The condition \eqref{eq:xi-3} combined with condition two of \eqref{eq:xi-12} ensures hyperbolicity of the system, while condition one of \eqref{eq:xi-12} ensures causality.

The conditions \eqref{eq:xi-12}, \eqref{eq:xi-3} came from demanding that the roots of the characteristic equation \eqref{ap2:char_eq} are such that the system is causal. One can re-write these conditions in terms of the quantity $V_\mu = \xi_\mu / |\bm{\xi}|$, noting that $|\bm{\xi}| \neq 0$ unless $\xi_\mu = 0$. Then the constraints become
\begin{equation}\label{ap2:cons_real}
    |\Re(V_0(\bm{\xi}))| \leq 1, \qquad \Im(V_0(\bm{\xi})) = 0, \qquad  \text{num}(V_0(\bm{\xi})) =  g \,.
\end{equation}
Now, let us consider the dispersion relations. Plane-waves $\delta U^B = \delta \tilde{U}^B(K) \exp[i K_\mu x^\mu]$, where $K_\mu = \{-\omega, \k\}$, solve the original equation \eqref{ap2:linearized} as long as
\begin{equation}\label{ap2:det_eq}
\begin{split}
     F(K_\mu) &\equiv \det\biggl[{\cal A}^{\mu_1 \mu_2 ... \mu_m}_{AB} (i)^m K_{\mu_1} K_{\mu_2} ... K_{\mu_m} \\
     &+ {\cal B}^{\mu_1 \mu_2 ... \mu_{(m-1)}}_{AB} (i)^{m-1}K_{\mu_1} K_{\mu_2} ... K_{\mu_{(m-1)}} + ... + C_{AB}\biggr]= 0 \,,
     \end{split}
\end{equation}
solving which gives rise to $\omega = \omega(\k)$. As shown in the main text, covariant stability\footnote{One could, instead, simply demand the constraints \eqref{eq:c-1}, \eqref{eq:c-2}, without the additional conditions of covariant stability; the linearity of $\omega(\k)$ then follows.} \eqref{eq:GFC} implies that $\omega(\k)$ is at most linear at large $\k$. Then one can define $V'_\mu = \lim_{\k\to\infty} K_\mu(\k)/|\k|$, which is finite. Dividing through \eqref{ap2:det_eq} by $|\k|^{g}$ (where, for equations of the form \eqref{ap2:linearized}, $g$ is simply $m \times n$) and taking the large-$\k$ limit yields
\begin{equation}
\label{eq:Finf}
    \det\biggl[{\cal A}^{\mu_1 \mu_2 ... \mu_m}_{AB}  V'_{\mu_1} V'_{\mu_2} ... V'_{\mu_m}\biggr]  = 0 \,,
\end{equation}
which is again the characteristic equation \eqref{ap2:char_eq}, now written in terms of $V'_\mu$. As discussed in the main text, covariant stability in classical theories implies that 
\begin{align}
  \lim_{|\k| \to \infty} \frac{|\Re\, \omega(\k)|}{|\k|} \leq 1, \qquad \lim_{|\k| \to \infty} \frac{\Im\, \omega(\k)}{|\k|} = 0, \qquad \text{num}(\omega) = {\cal O}_{|\k|}(F(a |\k|, {\bf s} |\k|)) = g \,,
\end{align}
which one can equivalently write as
\begin{equation}\label{ap2:cons_mom}
    |\Re (V'_0(\k))| \leq 1, \qquad  \Im ( V'_0(\k) ) = 0, \qquad \text{num}(V'_0(\k)) = {\cal O}_{|\k|}(F(a |\k|, {\bf s} |\k|)) = g \,.
\end{equation}
The constraints \eqref{ap2:cons_real} imposed on $V_\mu$ to render the theory causal from the point of view of characteristics are the same as the constraints \eqref{ap2:cons_mom} on the large-$\k$ dispersion relations. 
Therefore, demanding that the large-$\k$ dispersion relations obey the constraints \eqref{ap2:cons_mom} amounts to requiring that the theory is causal.


\section{Convergent Expansion}
\label{ap:ProofPowerSeries}
For an isotropic system, let us choose the wavevector $\k$ along $\x$, and define $k\equiv k_x$. Then the spectral curve of the system is a finite-order polynomial in $\omega$ and $k$. The polynomial is of order $m$ in $\omega$, and may generically be written in the form
\begin{equation}
\label{eq:Fwk-2}
    F(\omega,k) = a_m(k) \omega^m + a_{m-1}(k) \omega^{m-1} + ... + a_1(k) \omega + a_0(k) \,,
\end{equation}
where the various $a_n(k)$ are themselves polynomials in $k$ of order $\ell_n$. We are interested in the behaviour of the large-$k$ expansion of the eigenfrequencies of the system. In order to proceed, we define $v \equiv 1/k$, and aim to construct an expansion about $v=0$. If any of the $\ell_n$ are non-zero, then $F(\omega,k)$ diverges as $v \to 0$. Let us denote the largest of the $\ell_n$ by $\ell_F$, and define a new spectral curve $G(\omega,v) \equiv v^{\ell_F} F(\omega,1/v)$, so that $G(\omega, v=0)$ is a polynomial in $\omega$.

We are interested in solving $G(\omega,v)=0$, and expressing the solution as $\omega=\omega(v)$ in a neighborhood of $(\omega_0, v_0)$. 
For non-infinite $\omega_0$ and $v_0$, we have the following expansions. If the first derivative of $G$ with respect to $\omega$ at $(\omega_0, v_0)$ does not vanish, the analytic implicit function theorem gives the Taylor series expansion for $\omega(v)$ about $v=v_0$,
\begin{align}
   \frac{\partial G}{\partial\omega}(\omega_0, v_0) \neq 0 :  \quad  \omega(v) = \sum_{n=0}^\infty c_n (v-v_0)^n \,.
\end{align}
If the first $(p-1)$ derivatives of $G$ with respect to $\omega$ at $(\omega_0, v_0)$ vanish, but the $p$-th derivative does not, we have the Puiseux series expansion for $\omega(v)$ about $v=v_0$,
\begin{equation}
\label{eq:puiseux_implicit}
        \frac{\partial G}{\partial \omega}(\omega_0,v_0) = 0, \quad 
        \frac{\partial^2 G}{\partial \omega^2}(\omega_0,v_0) = 0, \dots \quad 
        \frac{\partial^p G}{\partial \omega^p}(\omega_0,v_0) \neq 0  : \quad 
        \omega(v) = \sum_{n=0}^\infty c_n \zeta^n \,,
\end{equation}
where $\zeta^s = (v-v_0)$, and where $s$ is a positive integer which is less than or equal to the integer $p \geq 2$. There may be multiple expansions each with their own $s_i$ such that $\sum s_i = p$. If $s>1$, then the Puiseux expansions in that branch will be related to one another, being of the form
\begin{equation}
\label{eq:branches}
    \omega_j(v) =  \sum_{n=0}^\infty c_n e^{2 \pi i n j/s}\zeta^n \,,
\end{equation}
where $j = \{0, 1, ..., s-1\}$.
We are interested in the behaviour of $\omega$ in the neighbourhood of $v_0 = 0$. However, for causal physical systems $\omega_0$ may not necessarily be finite: for example, for the wave equation with propagation speed $c_s$, we have $\omega(v) = \pm (c_s/v) \to\infty$, hence such $\omega(v)$ cannot be represented by a series of the form \eqref{eq:puiseux_implicit}. 

In order to handle expansions in causal physical systems such as the wave equation, we define the new variable $u \equiv \omega v^{q}$, where $q$ is an as-yet unspecified real number, and aim to construct an expansion for $u(v)$ about $v=0$ and a finite $u=u_0$. Expressed in terms of $u$, the spectral curve is
\begin{equation}\label{eq:G_rm}
    G(v^{-q} u, v) = v^{r_m} b_m(v) u^m + ... + v^{r_1} b_1(v) u + v^{r_0} b_0(v) \,,
\end{equation}
where $r_n \equiv \ell_F - \ell_n - n q$, and the $b_n(v) \equiv a_n(1/v)\, v^{\ell_n}$ are polynomials in $v$; we have $b_n(0) \neq 0$ and finite for all $n$.

Suppose one sets $q$ to be some sufficiently large number. Then each of the $v^{r_n}$ in \eqref{eq:G_rm} will have negative exponents, and $G(v^{-q} u, v)$ will diverge when $v = 0$. Similarly to how $G(\omega,v)$ was defined in the first place, we can define a new spectral curve which is a finite polynomial at $v=0$. Since for sufficiently large $q$, $r_m$ will be the most negative of the $r_n$, we can define 
\begin{equation}
    H_m(u,v) \equiv v^{-r_m} G(v^{-q} u, v) =  b_m(v) u^m + ... + v^{r_1{-}r_m} b_1(v) u + v^{r_0{-}r_m} b_0(v) \,.
\end{equation}
As $r_m$ is the most negative of the $r_n$ for sufficiently large $q$, $r_n - r_m \geq 0$ for all $n$, and thus
\begin{equation}
    H_m(u_0,0) = b_m(0) u_0^m = 0 \,.
\end{equation}
Therefore, there will be $m$ expansions $u=u(v)$ of the form \eqref{eq:puiseux_implicit} about the point $(u = 0, v=0)$, which is non-infinite. These $m$ expansions are convergent by the Puiseux theorem. We may now ask the follow-up question: for which values of $q$ are there non-zero $u_0 = u(v=0)$?

This question is relevant because $u_0 = u(v=0)$ is the first term of the expansion $u(v)$, and so upon transforming back to $\omega(v)$, we find that $\omega(v) = u_0 v^{-q} + ...\,$, where the dots refer to terms that are of higher-powers in $v$. In other words, $u_0$ is the coefficient of the highest-order term in the large-$k$ expansion of $\omega=\omega(k)$, and therefore the values of $q$ which yield non-zero $u_0$ are the respective orders in $k$ at which the large-$k$ expansions $\omega=\omega(k)$ begin. For example, the wave equation has $\omega = \pm c_s/v = \pm c_s k$, and one finds that $H_{m=2}(u_0,0)=0$ has non-zero solutions $u_0=\pm 1$ when $q=1$. 

One can use the method of Newton's polygon \cite{Wall-singular-points} to determine which values of $q$ lead to expansions of $u(v)$ with $u(0)\neq 0$ (as well as what the $s_i$ are for each branch, a feature we will not make use of here). To start with, by plotting the various linear functions $r_n$ of \eqref{eq:G_rm} against $q$, one finds that the only values of $q$ for which non-zero solutions $u_0 = u(v=0)$ exist are those for which the lines $r_n(q)$ intersect one another. 

It's quite straightforward to show that this must be the case. Consider a value $q=q_0$ where $r_n(q_0)$ are all different, i.e.\ the lines $r_n(q)$ do not intersect at $q_0$. Then there exists some $r_{n_1}$ which is the most negative of all the $r_n$ at $q_0$. Then we can define a new spectral curve
\begin{equation}
\label{eq:Hn1}
    H_{n_1}(u,v) \equiv v^{-r_{n_1}} G(v^{-q} u, v) = v^{r_{m}-r_{n_1}} b_{m}(v) u^v + ... + v^{r_1 - r_{n_1}} b_1(v) u + v^{r_0 - r_{n_1}} b_0(v).
\end{equation}
Since $r_{n_1}$ is the most negative, $r_{n} - r_{n_1}$ is positive for all $n$ except $n_1$. Therefore, 
\begin{equation}
    H_{n_1}(u_0,0) = b_{n_1}(0) u_0^{n_1} = 0 \,,
\end{equation}
which gives $u_0=0$; therefore, the expansion $\omega(v)$ can't start at a value of $q$ where an intersection doesn't occur. 

However, this doesn't necessarily mean that there are non-zero $u_0$ at every value of $q$ for which there is an intersection of the $r_n$. Suppose there are two $r_{n}$, call them $r_{n_2}(q)$ and $r_{n_3}(q)$, which intersect, but that at the intersection point $q_0$ both $r_{n_2}$ and $r_{n_3}$ are larger than the most negative $r_n$, which we once again label $r_{n_1}(q)$. Then we can once again define $H_{n_1}$ by \eqref{eq:Hn1}, and $r_{n_2} - r_{n_1} = r_{n_3} - r_{n-1} > 0$, and so we again find that
\begin{equation}
    H_{n_1}(u_0,0) = b_{n_1}(0) u_0^{n_1} = 0 \,.
\end{equation}
Therefore, the only values of $q$ for which there exist non-zero values of $u_0$ are those for which the most negative $r_n$, call it $r_{n_1}$, has an intersection. Since the $r_n(q)$ are linear functions of $q$, which of the $r_n$ is the most negative changes across the intersection. In other words, considering the envelope of the set of $r_n(q)$ from below, there exist non-zero values of $u_0$ whenever the slope of the envelope changes. In general, the number of non-zero solutions to $H_{n_1}(u_0,0) = 0$ at the intersection will be equal to the magnitude of the change of the slope of the envelope across the intersection. 

The intersection of two linear functions $r_{n_1}$ and $r_{n_2}$ with integer coefficients on the lower envelope occurs at
\begin{equation}
\label{eq:q_0_location}
    q_0 =  \frac{\ell_{n_1} - \ell_{n_2}}{n_2 - n_1} \,.
\end{equation}
We can similarly consider an additional line, $r_{n_3}$, which intersects at that point. Then
\begin{equation}
\label{eq:q_0_location_multi}
    q_0 = \frac{\ell_{n_1} - \ell_{n_2}}{n_2 - n_1} = \frac{\ell_{n_1} - \ell_{n_3}}{n_3 - n_1} = \frac{\ell_{n_2} - \ell_{n_3}}{n_3 - n_2} \,,
\end{equation}
and, in general, there can be an arbitrary number of lines intersecting at $q_0$. 
The analytic implicit function and Puiseux theorems guarantee the existence of convergent expansions of $\omega(k)$ in the $k\to\infty$ limit. Using the argument in the main body of the paper, one can see that covariant stability implies that the first term of the expansion is an integer power of $1/k$, hence $q$ is integer and $q\leq1$.

Finally, as an additional note, one can see from \eqref{eq:q_0_location_multi} that the only way for $q_0 = 0$ for all modes is if $\ell_{n}$ is the same for every term of $F(\omega,k)$. Since $\ell_{m}=0$ by \eqref{eq:c-2}, this implies that for any $F(\omega,k)$ with non-trivial $k$-dependence, there must be at least one mode with non-zero propagation speed $\lim_{k \to \infty} \omega/k$, i.e.\ at least one mode with $q_0 = 1$.

\begin{figure}
    \centering
    \includegraphics[scale=0.36]{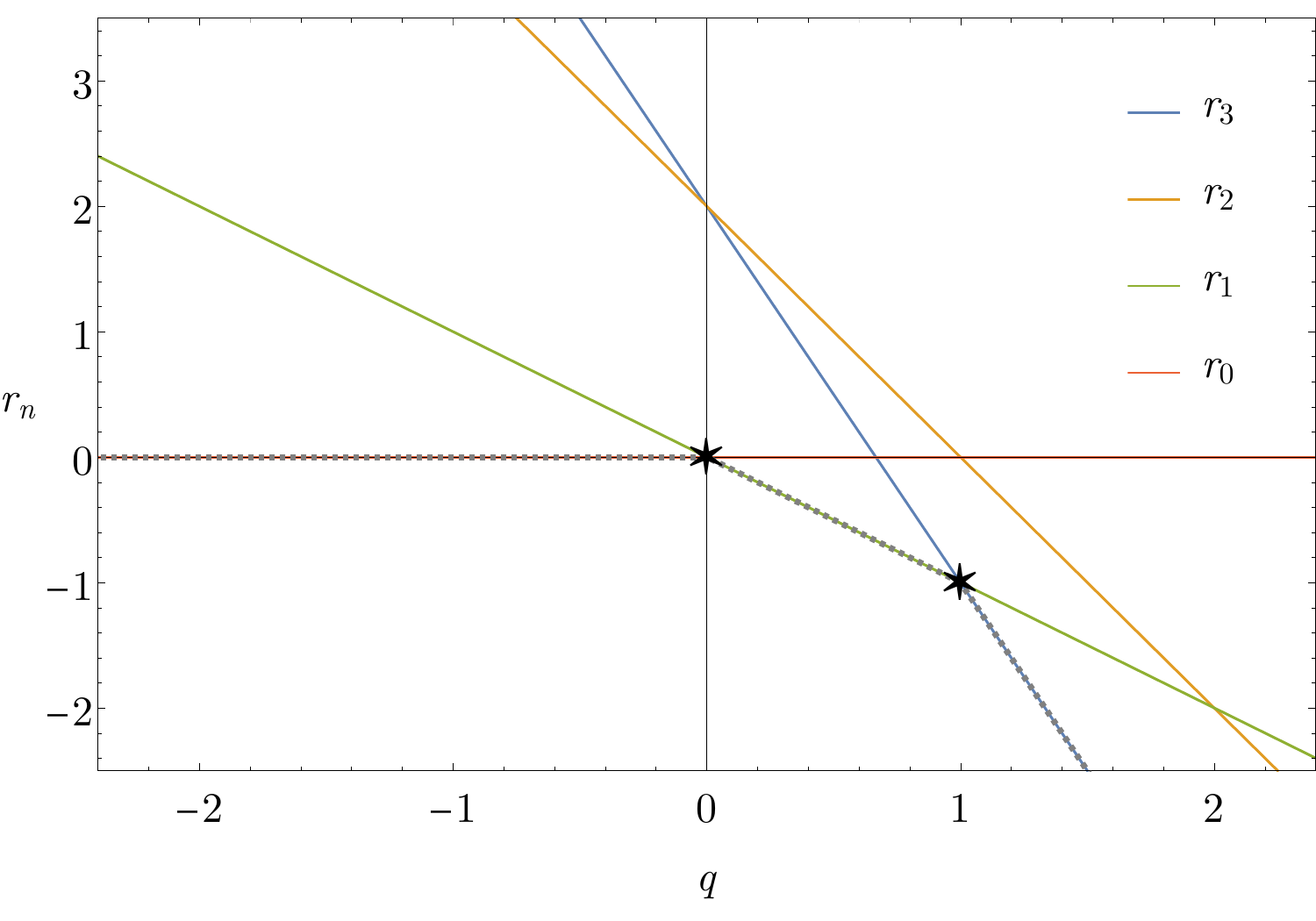}
    \caption{An example of plotting $r_n$ vs $q$ for the polynomial $\lr{\omega^2 - \frac{1}{2} k^2} (i \omega) - \omega^2 - i \omega + \frac{1}{4} k^2 $. The envelope of $r_n$ is indicated by a dotted line. For $q > 1$, $r_3$ is the smallest $r_n$. At $q=1$, there is an intersection. The slope changes from $-3$ to $-1$, indicating there are two modes which have $q=1$. Then, $r_1$ is the smallest until it intersects $r_0$ at $q=0$. The slope changes from $-1$ to $0$, and so there is one mode which has $q=0$. From there on, $r_0$ is the smallest.}
    \label{fig:rn_vs_q}
\end{figure}

To finish the appendix, we provide a brief example. Given the spectral curve 
\begin{equation}
    F(\omega,k) = \lr{\omega^2 - \frac{1}{2} k^2} (i \omega) - \omega^2 - i \omega + \frac{1}{4} k^2 ,
\end{equation} we can see that $G(\omega,v) \equiv v^{2} F(\omega,1/v)= i v^2 \omega^3 - v^2 \omega^2 - i v^2 \omega - \frac{i}{2} \omega + \frac{1}{4}$. We therefore find that
\begin{equation}
    G(v^{-q} u,v) = i v^{2-3q} u^3  - v^{2-2q} u^2 - \frac{1}{2} i v^{-q} \lr{1 + 2 v^2} u + \frac{1}{4} \,.
\end{equation}
We can read off from $G(v^{-q} u,v)$ that $r_3 = 2-3q$, $r_2 = 2-2q$, $r_1 = -q$, and $r_0 = 0$. For large values of $q$, the most negative $r_n$ will be $r_3$, and so $n_1 = 3$ for large $q$. If we plot these $r_n(q)$ against $q$ (as shown in Figure \ref{fig:rn_vs_q}), we will see that $r_3$ intersects $r_1$ at $q = q_0 = 1$. To the left of $q=1$, the most negative $r_n$ is $r_1$, and so the magnitude of the change of slope is two. We therefore expect to find two non-zero values for $u_0$ when $q=1$. 

Setting $q=1$, then, we can see that $r_3(q=1) =  -1$. Defining $H_3(u,v) = v G(v^{-1} u,v)$, we find that
\begin{equation}
    H_3(u,v) = i u^3 - v u^2 - \frac{1}{2} i \lr{1 + 2 v^2} u + \frac{v}{4} \,,
\end{equation}
and so
\begin{equation}
    H_3(u_0,0) =  i u_0  \lr{u_0^2 - \frac{1}{2}} = 0 \,,
\end{equation}
and therefore, as expected, we do indeed find two non-zero solutions for $u_0$, as well as one zero solution. These values of $u_0$ give the large-$k$ expansions $\omega(k) = \pm k/\sqrt{2} + O(1)$. Proceeding to lower the value of $q$, we see that $r_1$ (which is the most negative $r_n$ for $0<q<1$) intersects with $r_0$ at $q = 0$. Since the magnitude of the change of the slope of the envelope is one, we expect one solution for $u_0$ when $q=0$. Setting $q=0$ in $r_1$, we see that $r_1 = r_0 = 0$. Then $H_1(u,v) = G(u,v)$, and we find that
\begin{equation}
    H_1(u_0,0) = -\frac{i}{2} \lr{\frac{i}{2} + u_0} = 0 \,,
\end{equation}
which yields one non-zero solution, as expected. This value of $u_0$ gives the large-$k$ expansion $\omega(k) = -i/2 + O(1/k^2)$. The most negative $r_n$ for $q<0$ is $r_0$, which does not have any more intersections as $q \to -\infty$, and therefore the overall number of expansions is $2+1=3$, as expected for a system with $m=3$.

\bibliographystyle{JHEP}
\bibliography{hydro-general-biblio}

\end{document}